\documentclass{IEEEcsmag}

\IEEEoverridecommandlockouts

\usepackage[noadjust]{cite}
\usepackage{amsmath,amssymb,amsfonts}
\usepackage{algorithmic}
\usepackage{graphicx}
\usepackage{textcomp}
\usepackage{xcolor}
\usepackage{siunitx}
\usepackage{float}
\usepackage{stfloats}
\usepackage{todonotes}
\usepackage{tabularx}
\usepackage{hhline}
\usepackage{mathtools}
\usepackage{booktabs}
\usepackage[percent]{overpic}
\usepackage{subcaption}
\usepackage{tabu}
\usepackage{tikz}

\usetikzlibrary{shapes,arrows}
\usetikzlibrary{circuits.logic.US} 

\usepackage[utf8]{inputenc}
\usepackage{ulem}

\newcolumntype{Y}{>{\centering\arraybackslash}X}

\usepackage[acronym]{glossaries}

\def\BibTeX{{\rm B\kern-.05em{\sc i\kern-.025em b}\kern-.08em
    T\kern-.1667em\lower.7ex\hbox{E}\kern-.125emX}}
    
\newcommand{\rev}[1]{\textcolor{black}{#1}}
    
\newacronym{fc}{FC}{fabric controller}
\newacronym{gpio}{GPIO}{general-purpose input/output}

\newacronym[plural=FLLs,firstplural=Frequency-Locked Loops (FLLs)]{fll}{FLL}{Frequency-Locked Loop}
\newacronym{sne}{SNE}{Spiking Neural Engine}
\newacronym[plural=TCNs,firstplural=Temporal Convolutional Networks (TCNs)]{tcn}{TCN}{Temporal Convolutional Network}
\newacronym[plural=CNNs,firstplural=Convolutional Neural Networks]{cnn}{CNN}{Convolutional Neural Network}
\newacronym[plural=TNNs,firstplural=Ternarized Neural Networks]{tnn}{TNN}{Ternary Neural Network}
\newacronym{inq}{INQ}{Incremental Network Quantization}
\newacronym{ste}{STE}{Straight-Through-Estimator}
\newacronym{bn}{BN}{Batch Normalization}
\newacronym{dma}{DMA}{Direct Memory Access}
\newacronym{simd}{SIMD}{Single Instruction Multiple Data}
\newacronym{lr}{LR}{Learning Rate}
\newacronym[plural=PTUs, firstplural={Pan-Tilt Units}]{ptu}{PTU}{Pan-Tilt Unit}
\newacronym{mdf}{MDF}{Medium-density fibreboard}
\newacronym{cvat}{CVAT}{Computer Vision Annotation Tool}
\newacronym{coco}{COCO}{Common Objects in Context}
\newacronym{soa}{SoA}{State of the Art}
\newacronym{sf}{SF}{Sensor Fusion}
\newacronym{lorawan}{LoRaWAN}{Long Range Wide Area Network}
\newacronym{lora}{LoRa}{Long Range}
\newacronym{apb}{APB}{Advanced Peripheral Bus}
\newacronym{cdc}{CDC}{Clock Domain Crossing}

\newacronym[plural=DVS, firstplural={Dynamic Vision Sensors (DVS)}]{dvs}{DVS}{Dynamic Vision Sensor}
\newacronym[plural=FPGAs, firstplural={Field Programmable Gate Arrays (FPGAs)}]{fpga}{FPGA}{Field Programmable Gate Array}
\newacronym[plural=SNNs, firstplural={Spiking Neural Networks (SNNs)}]{snn}{SNN}{Spiking Neural Network}
\newacronym[plural=DNNs, firstplural={Deep Neural Networks (DNNS)}]{dnn}{DNN}{Deep Neural Network}

\newacronym{fbk}{FBK}{Fondazione Bruno Kessler}
\newacronym{FGSM}{FBK}{Fast Gradient Sign Method}
\newacronym{date}{DATE}{Design Automation and Test in Europe}
\newacronym{iimtc}{I2MTC}{The International Instrumentation \& Measurement Technology Conference}
\newacronym{ini}{INI}{Institute of Neuroinformatics}

\newacronym[plural=LUTs, firstplural={Lookup Tables (LUTs)}]{lut}{LUT}{Lookup Table}

\newacronym{lpwan}{LPWAN}{Low-Power Wide Area Network}
\newacronym{nbiot}{NB-IoT}{Narrow Band Internet-of-Things}

\newacronym{saer}{SAER}{Synchronous Address-Event Representation}
\newacronym{fps}{FPS}{Frames Per Second}
\newacronym{vcd}{VCD}{Value-Change Dump}
\newacronym{spi}{SPI}{Serial Peripheral Interface}
\newacronym{cpi}{CPI}{Camera Parallel Interface}
\newacronym{fifo}{FIFO}{First-In First-Out Queue}
\newacronym{mcu}{MCU}{Microcontroller}
\newacronym{ble}{BLE}{Bluetooth Low-Energy}
\newacronym{uart}{UART}{Universal Asynchronous Receiver-Transmitter}
\newacronym{sta}{STA}{Static Timing Analysis}
\newacronym{ptz}{PTZ}{Pan-Tilt Unit}

\newacronym{CV}{CV}{Computer Vision}
\newacronym{EoT}{EoT}{Expectation over Transformation}
\newacronym{RPN}{RPN}{Region Proposal Network}
\newacronym{TV}{TV}{Total Variation}
\newacronym{NPS}{NPS}{Non-Printability Score}
\newacronym{STN}{STN}{Spatial Transformer Network}
\newacronym{MTCNN}{MTCNN}{Multi-Task Convolutional Neural Network}
\newacronym{YOLO}{YOLO}{You Only Look Once}
\newacronym{SSD}{SSD}{Single Shot Detector}
\newacronym{SOTA}{SOTA}{State of the Art}
\newacronym{NMS}{NMS}{Non-Maximum Suppression}
\newacronym{ic}{IC}{Integrated Circuit}
\newacronym{rf}{RF}{Radio Frequency}
\newacronym{tcxo}{TCXO}{Temperature Controlled Crystal Oscillator}
\newacronym{jtag}{JTAG}{Joint Test Action Group industry standard}
\newacronym{swd}{SWD}{Serial Wire Debug}
\newacronym{sdio}{SDIO}{Serial Data Input Output}
\newacronym{ldo}{LDO}{Linear Dropout Regulator}

\newacronym[plural=PCBs, firstplural={Printed Circuit Boards (PCB)}]{pcb}{PCB}{Printed Circuit Board}
\newacronym[plural=ASICs, firstplural={Application Specific Integrated Circuits}]{asic}{ASIC}{Application Specific Integrated Circuit}

\newacronym[plural=BNNs, firstplural={Binary Neural Networks (BNNs)}]{bnn}{BNN}{Binary Neural Network}
\newacronym[plural=NNs, firstplural={Neural Networks}]{nn}{NN}{Neural Network (NNs)}
\newacronym[plural=SCMs, firstplural={Standard Cell Memories (SCMs)}]{scm}{SCM}{Standard Cell Memory}
\newacronym{ann}{ANN}{Artificial Neural Networks}
\newacronym{ml}{ML}{Machine Learning}
\newacronym{ai}{AI}{Artificial Intelligence}
\newacronym{iot}{IoT}{Internet of Things}
\newacronym{fft}{FFT}{Fast Fourier Transform}
\newacronym[plural=OCUs, firstplural={Output Channel Compute Units (OCUs)}]{ocu}{OCU}{Output Channel Compute Unit}
\newacronym{alu}{ALU}{Arithmetic Logic Unit}
\newacronym{mac}{MAC}{Multiply-Accumulate}
\newacronym{soc}{SoC}{System-on-Chip}

\newacronym{PGD}{PGD}{Projected Gradient Descend}
\newacronym{CW}{CW}{Carlini-Wagner}
\newacronym{OD}{OD}{Object Detection}

\newacronym{rrf}{RRF}{RADAR Repetition Frequency}
\newacronym{nlp}{NLP}{Natural Language Processing}
\newacronym{qam}{QAM}{Quadrature Amplitude Modulation}
\newacronym{rri}{RRI}{RADAR Repetition Interval}
\newacronym{radar}{RADAR}{Radio Detection and Ranging}
\newacronym{loocv}{LOOCV}{Leave-one-out cross validation}
\newacronym{raw}{RAW}{Read-After-Write}
\newacronym[plural=ISAs, firstplural={Instruction Set Architectures (ISAs)}]{isa}{ISA}{Instruction Set Architecture}
\newacronym{mhsa}{MHSA}{Multi-Head Self-Attention}
\newacronym{os}{OS}{Operating System}
\newacronym{bsp}{BSP}{Board Support Package}
\newacronym{ttn}{TTN}{The Things Network}
\newacronym{wip}{WIP}{Work in Progress}

\newacronym{cls}{CLS}{Classification Error}
\newacronym{loc}{LOC}{Localization Error}
\newacronym{bkgd}{BKGD}{Background Error}

\newacronym{dsp}{DSP}{Digital Signal Processing}

\begin{document}

\sptitle{Department: Head}
\editor{Editor: Name, xxxx@email}

\title{
TCN-CUTIE: A 1036~TOp/s/W, 2.72~µJ/Inference, 12.2~mW All-Digital Ternary Accelerator in 22~nm FDX Technology
}

\author{Moritz Scherer\IEEEauthorrefmark{2} Alfio Di Mauro\IEEEauthorrefmark{2}, Tim Fischer\IEEEauthorrefmark{2}, Georg Rutishauser\IEEEauthorrefmark{2}, Luca Benini\IEEEauthorrefmark{2}\IEEEauthorrefmark{3}}
\affil{\IEEEauthorrefmark{2}Dept. of Information Technology and Electrical Engineering, ETH Z\"{u}rich, Switzerland\\
\IEEEauthorrefmark{3}Dept. of Electrical, Electronic and Information Engineering, University of Bologna, Italy}

\begin{abstract}
Tiny Machine Learning (TinyML) applications impose µJ/Inference constraints, with a maximum power consumption of tens of mW. It is extremely challenging to meet these requirements at a reasonable accuracy level. This work addresses the challenge with a flexible, fully digital Ternary Neural Network (TNN) accelerator in a RISC-V-based System-on-Chip (SoC). Besides supporting Ternary Convolutional Neural Networks, we introduce extensions to the accelerator design that enable the processing of time-dilated Temporal Convolutional Neural Networks (TCNs).
The design achieves 5.5 µJ/Inference, 12.2 mW, 8000 Inferences/sec at 0.5 V for a Dynamic Vision Sensor (DVS) based TCN, and an accuracy of 94.5\% and  2.72 µJ/Inference, 12.2 mW, 3200 Inferences/sec at 0.5 V for a non-trivial 9-layer, 96 channels-per-layer convolutional network with CIFAR-10 accuracy of 86\%. The peak energy efficiency is 1036 TOp/s/W, outperforming the state-of-the-art silicon-proven TinyML quantized accelerators by 1.67x while achieving competitive accuracy.


\end{abstract}

\maketitle

\section{Introduction}
Advances in \gls{ml} research in recent years have enabled a new direction of research within the field of embedded systems, called TinyML, targeting the execution of non-trivial ML tasks within the strict constraints of low-power (mW) embedded devices.
TinyML is becoming increasingly pervasive with applications including wearable computer vision, gesture recognition, and many more. The key challenges in TinyML are energy efficiency at a few mW of power while ensuring accurate and fast inference. Specialized TinyML accelerators tackle both challenges by providing high throughput at low power, but often they do so by compromising accuracy or by specializing on a single network topology, with no flexibility. 
We present a flexible and accurate TinyML architecture, integrating CUTIE, a highly configurable \gls{tnn} accelerator based on a fully unrolled compute architecture \cite{scherer_cutie_2022} within the \textit{Kraken} RISC-V \gls{soc}. 
Further, we introduce a novel \gls{tcn} extension within the CUTIE architecture, enabling the execution of autonomous data-to-label time series prediction, a very common use case for TinyML devices. We present the deployment and power measurements of a neural network architecture exploiting these extensions, which achieves an accuracy of 94.5\% at an energy cost of \SI{5.5}{\micro\joule} per inference, performing gesture recognition from a \gls{dvs} trained on the DVS 128 dataset \cite{amir_low_2017}.
To the best of our knowledge, we are the first to demonstrate a peak core energy efficiency beyond \SI{1}{PetaOp\per\second\per\watt} for neural network inference in an all-digital and flexible platform, measured on a fabricated \gls{soc}.

\section{SoC Implementation}

\begin{figure*}[ht]
  \begin{center}
    \includegraphics[width=\linewidth]{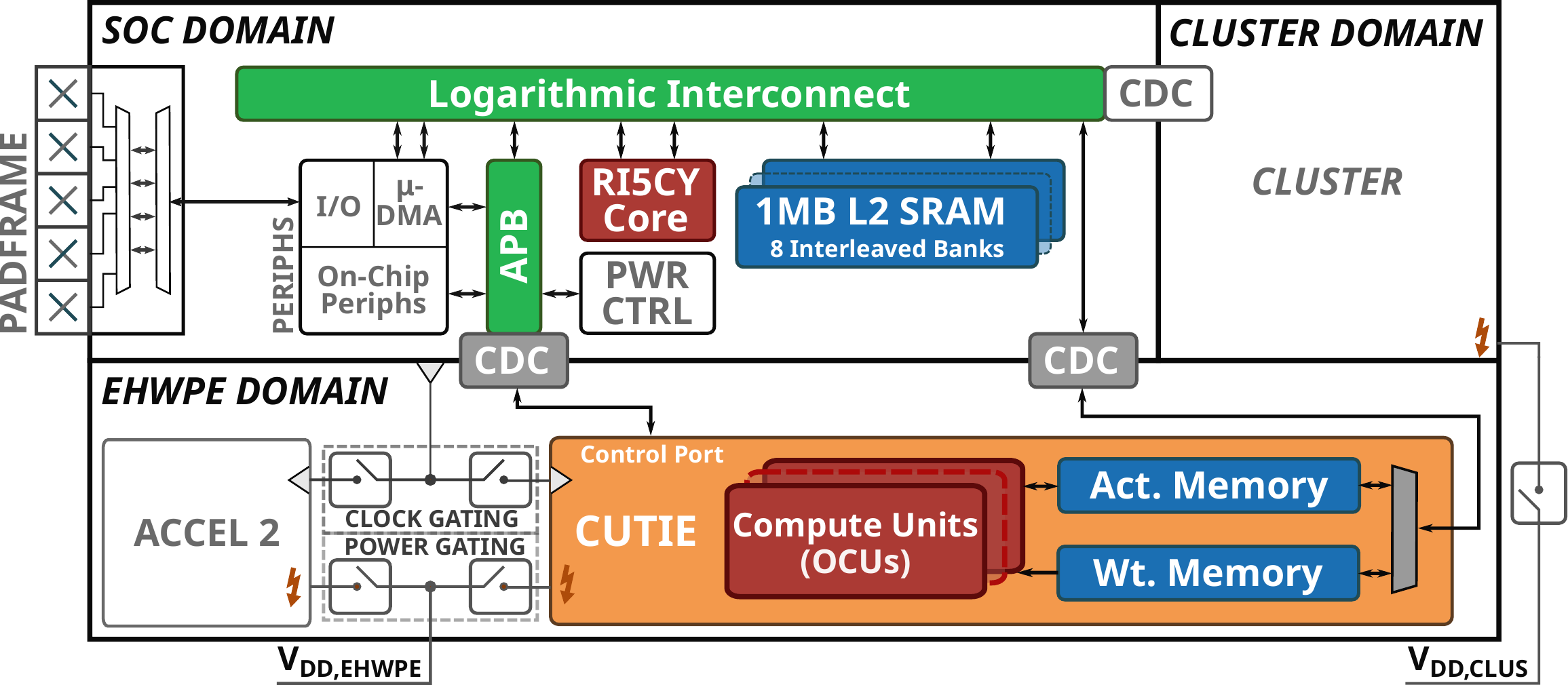}
    \caption{Block diagram of Kraken, including the three switchable power domains and always-on SoC domain. The CUTIE accelerator is integrated with a control port connected to the APB and a data port connected to the high-bandwidth logarithmic interconnect. The Cluster and Accel~2 IPs are not discussed in this paper.}
    \label{fig:toplevel}
  \end{center}
\end{figure*}

The Kraken \gls{soc} is a RISC-V-based microcontroller based on the Pulpissimo \gls{soc} \cite{schiavone_quentin_2018}. A RI5CY core \cite{gautschi_near-threshold_2017} serves as the \gls{fc}, coordinating the operation of the other subsystems. 
For parallel signal processing tasks, it contains an 8-core PULP cluster of RISC-V cores.
Kraken has an extensive set of peripherals for off-chip communication. They are implemented as {\textmu}DMA \cite{pullini_dma_2017} extensions, freeing the \gls{fc} from most management duties. On-chip peripherals include an event unit for interrupt mapping, a RISC-V-compliant debug unit for JTAG control of the chip and a power controller. 4 \gls{fll} modules provide independently run-time configurable clocks to the {\textmu}DMA peripherals, the \gls{soc} domain, the accelerator (EHWPE) domain, and the PULP cluster. 
Kraken has three core supply rails and four core power domains. The \gls{soc}, cluster and EHWPE domains each have a separate supply. Kraken features task-specific accelerators: In the following sections, we focus on the \gls{tnn} inference engine, its integration, and silicon measurements. 
The accelerators share their supply voltage, but each is located in its power domain and can be power-gated individually to minimize current draw by idle system components. 
A block diagram of the \gls{soc} architecture is shown in Figure~\ref{fig:toplevel}.


\section{CUTIE Design}

The Completely Unrolled Ternary Inference Engine (CUTIE), is a highly configurable \gls{cnn} accelerator architecture for completely ternarized neural networks, introduced in \cite{scherer_cutie_2022}. In contrast to systolic arrays, CUTIE uses a completely unrolled compute architecture, which means that one \gls{ocu} is allocated for every output channel, making the computation output-stationary. Further, each \gls{ocu} includes weight buffers, minimizing weight data movement. Each \gls{ocu} processes a full activation window per cycle, without pipelining in the compute units, making the architecture also input-stationary. 
To fully exploit all opportunities for data reuse in \glspl{cnn}, a linebuffer designed to eliminate data access stalling is added. Thanks to this highly parallel design, CUTIE fully exploits data reuse at all levels and minimizes data movement. In addition, ternary weights and activations enable the exploitation of zero values to translate sparsity into reduced toggling in the compute units. Thus, minimized data movement and switching activity are the cornerstones of CUTIE's efficiency.

In this work, we extend the CUTIE \gls{tnn} accelerator to support hybrid 2D-\gls{cnn} \& 1D-\gls{tcn} networks. \rev{As demonstrated in \cite{rutishauser_ternarized_2022}, the combination of low precision, i.e. ternarized, CNN and TCN achieved superior accuracy in classifying time-distributed data like streams of events produced by event-based sensors, like DVS cameras. Data produced by such sensors are characterized by a high level of unstructured sparsity and exhibit both short and long temporal dynamics. A hybrid CNN-TCN approach allows fine-tuning the network capabilities to achieve the highest accuracy when processing event streams. Specifically, the CNN captures the spatial dependency among neighboring events, that cluster in specific regions of the input feature map, as well as short temporal dependency among events belonging to consecutive time steps; an event happening in the scene tends to persist over multiple time steps. The 1D TCN extracts longer temporal dependencies among features distributed across the entire sample time window.} 1D-\glspl{tcn} use dilated convolutions \cite{lea_temporal_2016}, meaning feature map data is accessed in a strided fashion.
The extensions required to support 1D-\glspl{tcn} efficiently are twofold: 1) We designed a \gls{tcn} memory, enabling dilated feature map data access without stalling, 2) We implemented a scheduling algorithm that maps 1D dilated convolutions to 2D undilated convolutions, which make use of CUTIE's efficient compute architecture.

\section{TCN Extensions}

\begin{figure*}[ht]
  \begin{center}
    \includegraphics[width=\linewidth]{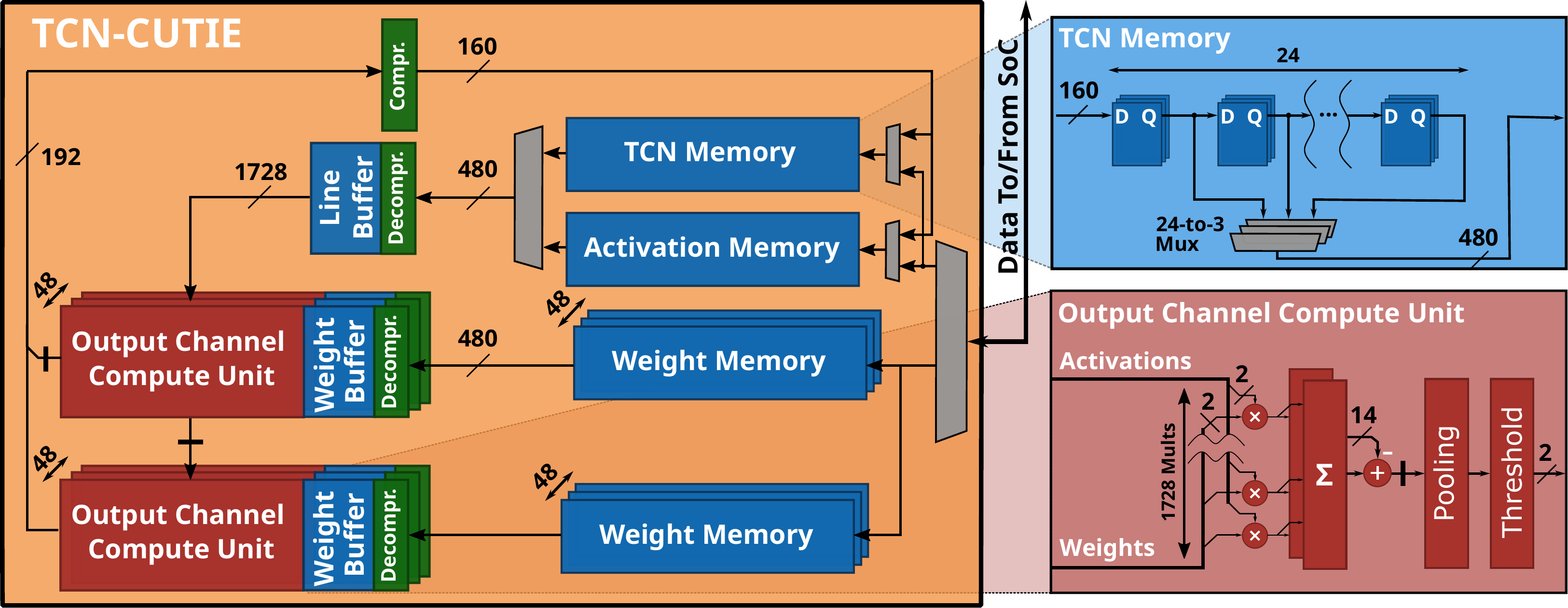}
    \caption{Block diagram of the 96-channel CUTIE implementation with TCN extensions, showing the completely unrolled data path. The insets show the flip-flop based TCN memory and the OCU, which processes an entire convolution window per cycle. Notably, the OCU uses a single pipeline stage.}
    \label{fig:cutie}
  \end{center}
\end{figure*}

To support hybrid 2D-\gls{cnn} \& 1D-\gls{tcn} networks, CUTIE has to be extended with a small memory, the \gls{tcn} memory, that can hold the 1D feature vectors that are extracted by each inference of a 2D-\gls{cnn}.
The \gls{tcn} memory enables the execution of hybrid 2D-CNN \& 1D-TCN networks, as well as pure 1D processing. The output of the \gls{tcn} memory has the same size as the activation memory, which is achieved by multiplexing three time steps according to the address of the first required pixel. In the Kraken \gls{soc}, the \gls{tcn} memory was dimensioned to hold a total of 24 feature vectors\rev{, corresponding to a memory size of only 576 bytes. Nevertheless, 24 time steps are sufficient to cover a long receptive window even at high framerates: if the 2D \gls{cnn} takes as input 15 stacked frames captured at a rate of 300 FPS ($5-10\times$ the speed of most ordinary cameras), the resulting receptive time window for a \gls{tcn} covering 24 time steps is still \SI{1.2}{\second}. Due to its small size, we implemented the \gls{tcn} memory/ as a flip-flop-based shift register to reduce leakage power.} A block diagram of the CUTIE \gls{tnn} accelerator with the proposed \gls{tcn} memory extension is shown in Figure~\ref{fig:cutie}. 

The second extension we introduce to the CUTIE accelerator is the mapping of 1D dilated convolutions. Dilated 1D convolutions with a kernel length $N$ and dilation factor $D$ of an input $x$ with a kernel $w$ can be described by their mathematical definition, shown in Equation~\ref{eq:tcn}: 
\begin{equation}\label{eq:tcn}
(w \star x)[n] = \sum_{k = 1}^{N} \tilde{x}[n-(k-1) \cdot D] \cdot w[N-k]
\end{equation}
where 
\begin{equation*}
    \tilde x[n] = \\
    \begin{cases}
    x[n], & n \ge 0\\
    0,  & else
    \end{cases}
\end{equation*}
is the \textit{causally padded} input vector x. \rev{The main advantage of dilated convolutions over undilated ones lies in their ability to reach a longer receptive field in fewer layers. In a \gls{tcn} with $N=3$ and $D_i=2^i$, where $D_i$ denotes the $i$-th layer's kernel dilation, the receptive field $f_k$ in layer $k$ can be calculated as $$f_k = 1 + \sum_{i=0}^{k}(N-1)\times 2^i$$. The receptive field increases exponentially with the number of layers, decreasing the number of layers needed to cover a given number of input steps. For the 24 input steps supported by TCN-CUTIE, the number of layers is reduced from 12 for undilated convolutions to 5 with exponentially increasing dilations. In a direct implementation} , the elements of $\tilde{x}$ are not accessed contiguously, instead, they are accessed with a stride of $D$. Due to the specialized memory hierarchy of CUTIE, non-contiguous or strided accesses lead to stalling, decreasing efficiency. To avoid this, we reformulate equation~\ref{eq:tcn} as a 2D correlation:
\begin{equation*}\label{eq:tcn_2d}
(w \star x)[n] = \sum_{k = 1}^{N} z[N-k, mod(n,D)] \cdot w[N-k]
\end{equation*}
where
\begin{equation*}
    z[n, m] = \tilde{x}[n \cdot D + m]
\end{equation*}

\begin{figure}[t]
  \begin{center}
    \includegraphics[width=\linewidth]{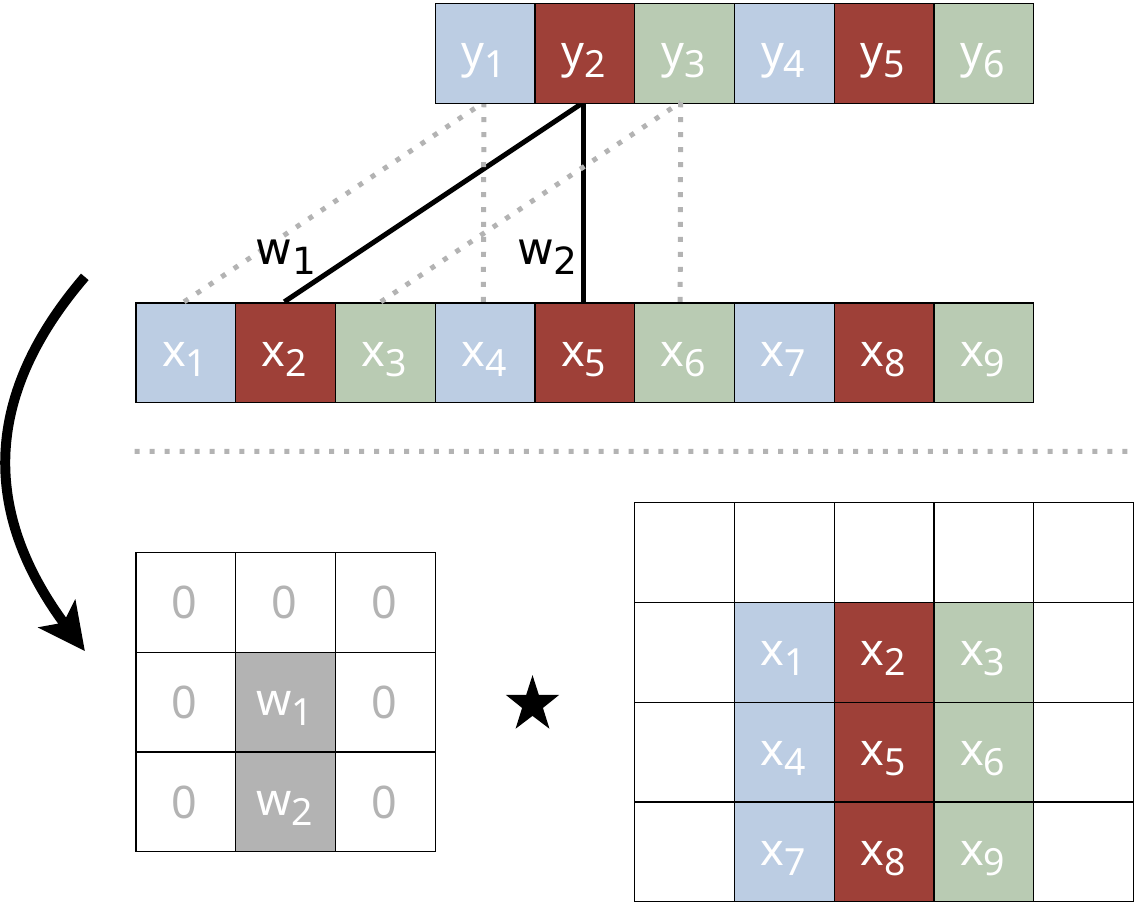}
    \caption{Example mapping of a dilated 1D convolution to an undilated 2D convolution for $D=3$, $N=2$}
    \label{fig:tcn}
  \end{center}
\end{figure}

\rev{A visual representation of this mapping is shown in Figure~\ref{fig:tcn}. To form the dense 2D feature map, the 1D vector is wrapped around after D elements. Further, zero padding (shown in white in Figure~\ref{fig:tcn}) is applied on the edges to implement the causality required by \glspl{tcn} as well as the correct start- and endpoint of the convolution. To respect the hardware constraints of CUTIE, i.e. weight kernels having size 3×3, the 1D weight kernel is projected into the middle column of the 2D weight kernel, while all other elements in the weight kernels are set to zero. This mapping ensures that the kernel dot product is only computed over a single column and the column elements are dilated by the dilation factor D. Since this mapping is fully equivalent to a 2D convolutional layer and all transforms necessary can be computed offline and require no data marshalling, it fully and efficiently reuses the CUTIE architecture with minor hardware overhead.}




\section{TCN-CUTIE Implementation}


Thanks to its highly configurable nature, the CUTIE architecture can be adapted to many application scenarios. In the Kraken \gls{soc}, we dimensioned the memories for feature map sizes of up to 64$\times$64 pixels with up to 96 channels. We designed the \gls{tcn} memory to hold a total of 24 time steps.
Since CUTIE's throughput per cycle is enormous due to the high degree of parallelism, we used relaxed timing constraints during synthesis, to enable extensive instantiation of low-leakage library cells. The CUTIE \gls{tnn} accelerator's clock can be hierarchically clock-gated to minimize idle switching activity in idle \glspl{ocu} when network layers have a small number of output channels. Inference can be triggered via a configuration register or an interrupt line from I/O peripherals, enabling autonomous data preparation and inference without intervention from the \gls{fc}. After inference has concluded, CUTIE asserts an interrupt which is used to wake up the \gls{fc}. 

\section{Kraken Physical implementation}

\begin{figure}[htb]
  \begin{center}
    \includegraphics[width=\linewidth]{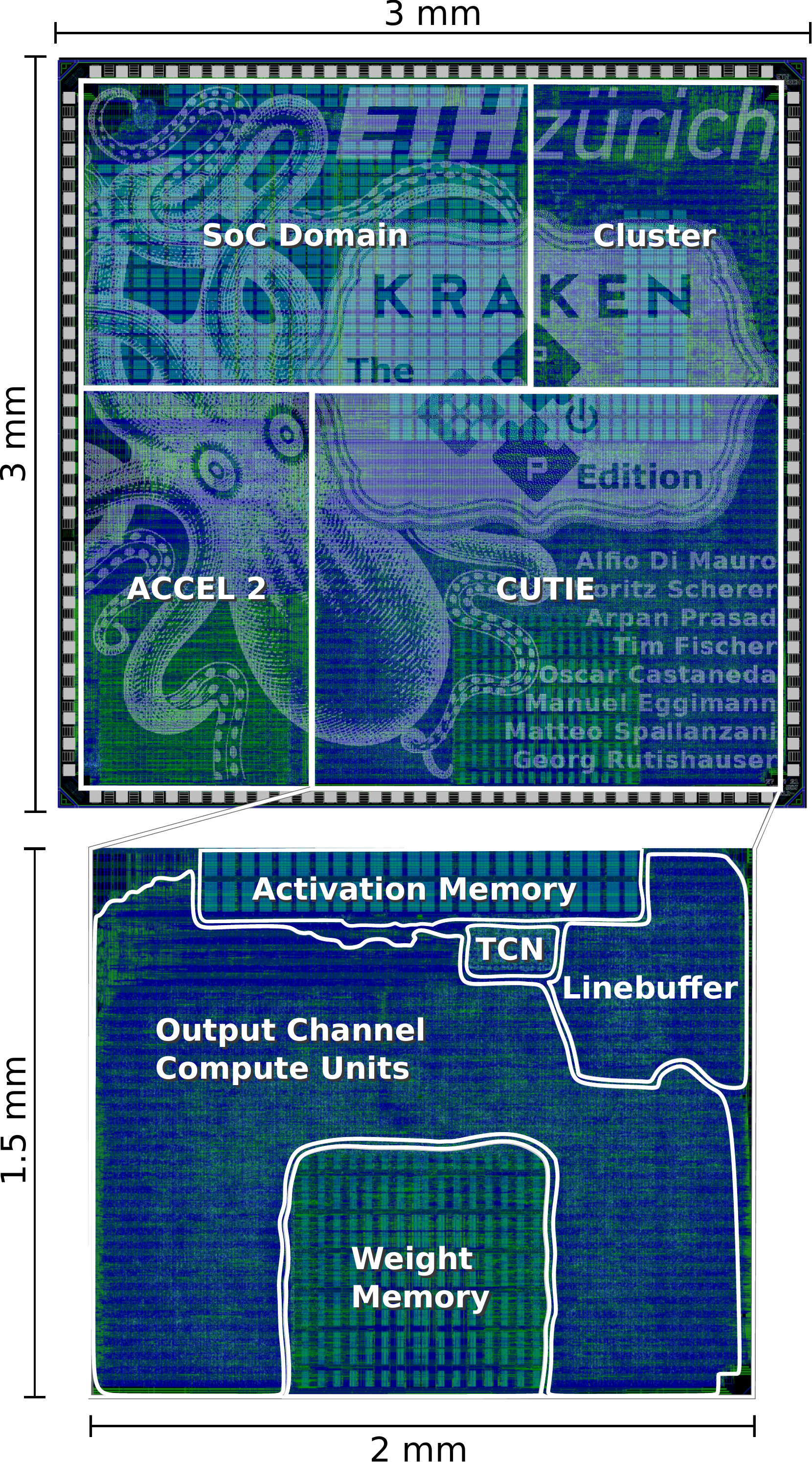}
    \caption{Die micrograph of the Kraken SoC. The top floorplan shows the four power domains, including SoC, Cluster, Accel 2 and CUTIE. The bottom floorplan shows the layout of modules within CUTIE.}
    \label{fig:dieshot}
  \end{center}
\end{figure}

\rev{The Kraken chip has been designed and manufactured in \textit{GlobalFoundries} 22nm technology, the total die area is \SI{9}{\milli \meter \squared}. The three Kraken subsystems are implemented as independent clock and power domains. Both the general-purpose RISC-V-based accelerator and the EHWPE domain can be entirely power-gated to reduce their leakage consumption when not in use. The chip can operate in a wide supply voltage range, i.e., from \SIrange{0.5}{0.9}{\volt}.
The chip host a total of 88 pads, 46 of which can be used either as GPIO or as an alternate function, i.e., as one of the signals of each IO peripherals, in an all-to-all muxing scheme. Figure~\ref{fig:dieshot} shows an annotated floorplan of the Kraken \gls{soc}, including the SoC Domain, Cluster, Accelerator 2, as well as CUTIE. The CUTIE accelerator occupies \SI{2.96}{\milli \meter \squared} of area. In the CUTIE layout, the area occupied by memory macros composing the internal buffers, and digital logic are highlighted.}
\rev{The memories including weight buffers in the OCUs take up 60\% of the total die area of CUTIE, while the rest is used by the compute units. The additional \gls{tcn} memory which holds the sequence samples was implemented in \gls{scm} and has a negligible impact of less than 1\% on area.}
\begin{figure}[ht]
  \begin{center}
    \includegraphics[width=\linewidth]{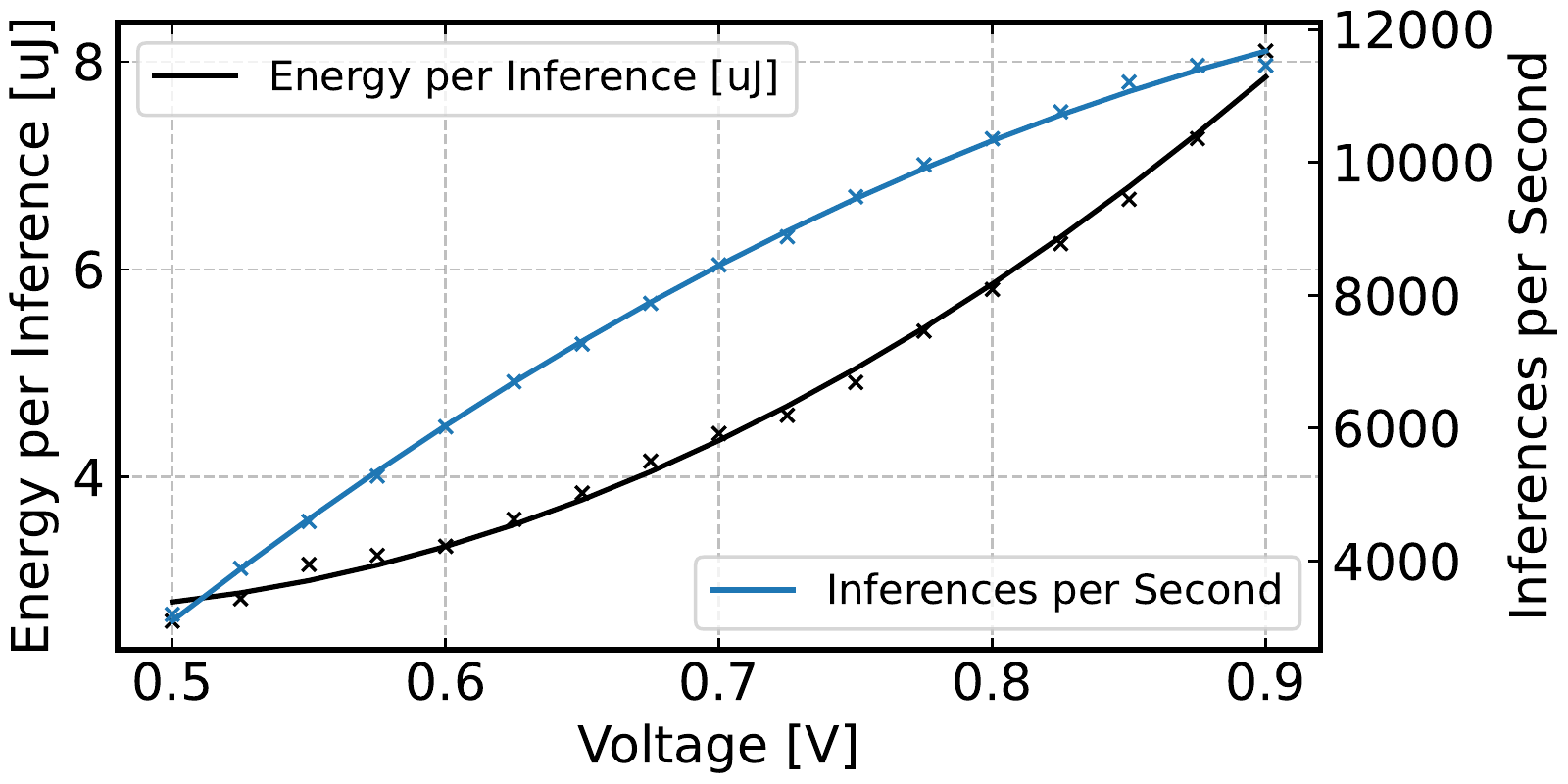}
    \includegraphics[width=\linewidth]{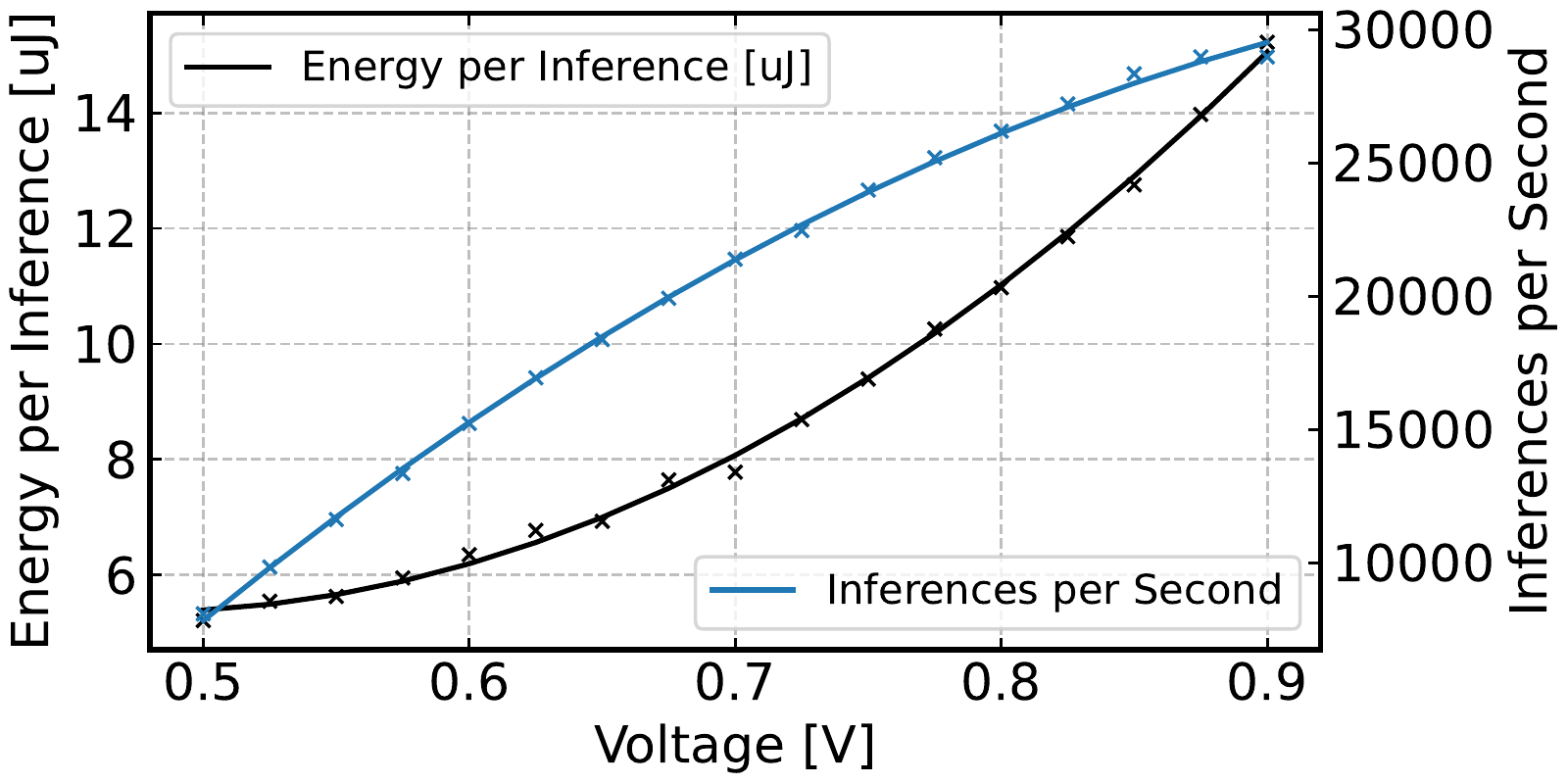}
    \caption{Energy per inference and inferences per second for the CIFAR-10 (upper) and DVS (lower) networks plotted against voltage using the maximum stable frequency at each corner. All data was recorded at 25°C.}
    \label{fig:fullnetwork}
  \end{center}
\end{figure}

\section{Evaluation}

To benchmark the accelerator's performance against similar state-of-the-art designs, we measure the execution of a ternarized 9-layer (8 CONV layers, 1 FC classifier) CIFAR-10 network as used in \cite{scherer_cutie_2022, knag_617_2020, moons_binareye_2018} with 96 instead of 128 channels. This network achieves an accuracy of 86\% on CIFAR-10, which is on par with the binarized version using 128 layers used in \cite{moons_binareye_2018, knag_617_2020}. 
Similarly, we execute the hybrid 2D-\gls{cnn} \& 1D-\gls{tcn} network proposed in \cite{rutishauser_ternarized_2022}, consisting of 5 2D-\gls{cnn} layers and 4 1D-\gls{tcn} layers that process 5 time steps. This network achieves an accuracy of 94.5\% on the 12-class DVS 128 dataset.

\begin{figure}[t]
  \begin{center}
    \includegraphics[width=\linewidth]{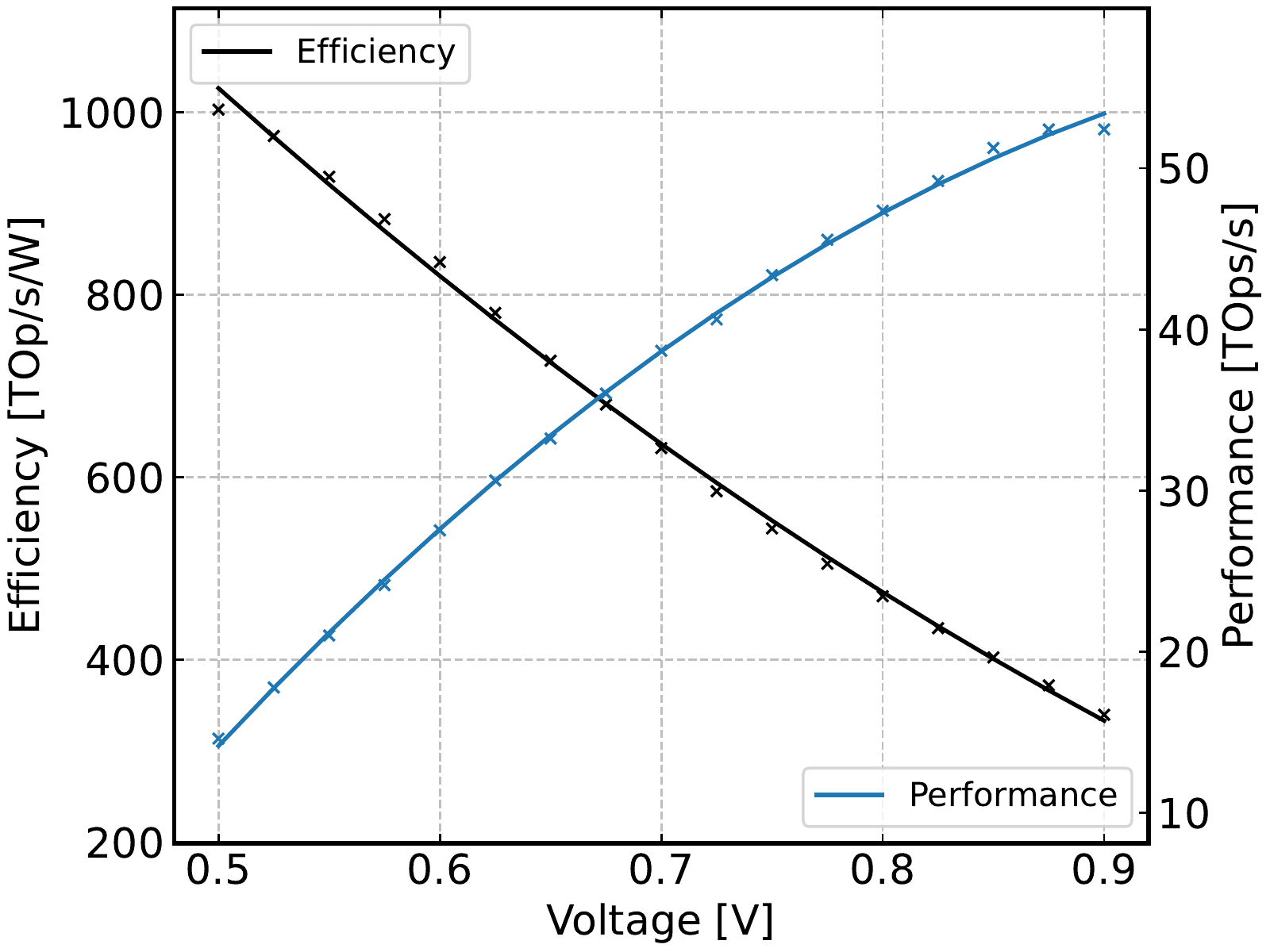}
    \caption{Peak energy efficiency and throughput plotted against voltage. One MAC operation corresponds to 2 Ops. Data is calculated at the maximal frequency for each voltage corner. All data was recorded at 25°C.}
    \label{fig:energyefficiency}
  \end{center}
\end{figure}

To evaluate the power consumption of CUTIE, we measured the current drawn by the Kraken ASIC on an ASIC tester, while running the deployed networks on CUTIE using pre-selected inputs which were randomly drawn from the respective validation set of the datasets used for training.
The presented power consumption numbers of the CUTIE accelerator include its memories, but do not include chip I/O energy. All measurements were performed at room temperature. 
\rev{We profiled the accelerator's performance at 25C over a range of \SI{0.5}{\volt} - \SI{0.9}{\volt}. Below \SI{0.5}{\volt}, the integrated SRAM macros start exhibiting bit errors. In terms of efficiency, we find that the \SI{0.5}{\volt} operating corner, operating at \SI{54}{\mega\hertz} achieves the lowest energy per inference of \SI{2.72}{\micro\joule} and \SI{5.5}{\micro\joule} at an average throughput of \SI{5.4}{TOp\per\second} and \SI{1.2}{TOp\per\second} for the CIFAR-10 and DVS networks, with a peak energy efficiency in the first layer of the CIFAR-10 network of \SI{1036}{TOp \per\second\per\watt} and peak throughput of \SI{14.9}{TOp \per\second}. The operating corner using \SI{0.9}{\volt} achieves the highest peak throughput of \SI{51.7}{TOp\per\second}, but a lower peak energy efficiency of \SI{318}{TOp\per\second\per\watt}. Figure~\ref{fig:fullnetwork} shows plots for throughput and energy per inference against voltage for the CIFAR-10 and DVS networks. Figure~\ref{fig:energyefficiency} shows the peak energy efficiency per operation and throughput versus voltage for the first layer of the CIFAR-10 network.}

\section{Comparison with State-of-the-Art}

\begin{table*}[bh]
\caption{Comparison of CUTIE with SoA highly quantized digital accelerators. All listed papers use the CIFAR-10 dataset and 9-layer CNN, however, this work uses 96 channels instead of 128.}  

  \begin{tabu}{|[1.25pt]X|r| r|r r@{\hspace{-0.2cm}}| r|r r@{\hspace{-0.2cm}} |[1.25pt]}
  \tabucline[1.25pt]{-}
    Characteristics & \cite{moons_binareye_2018} &  \multicolumn{2}{r}{\cite{knag_617_2020}}& & \multicolumn{2}{r}{\textbf{This work}} &\\ \tabucline[1.25pt]{-}
    \hline
    Computation Method & digital & \multicolumn{2}{r}{digital}& & \multicolumn{2}{r}{digital} & \\
    \hline
    Weight Precision  & binary & \multicolumn{2}{r}{binary}& & \multicolumn{2}{r}{ternary}& \\
    \hline
    Activation Precision & binary & \multicolumn{2}{r}{binary}& & \multicolumn{2}{r}{ternary} & \\
    \hline
    Technology & \SI{28}{\nano\meter} & \multicolumn{2}{r}{\SI{10}{\nano\meter}}& &  \multicolumn{2}{r}{\SI{22}{\nano\meter}} & \\
    \hline
    Dataset & CIFAR-10 & \multicolumn{2}{r}{CIFAR-10}& & \multicolumn{2}{r}{CIFAR-10} & \\
    \hline
    Accuracy & 86\% & \multicolumn{2}{r}{86\%}& & \multicolumn{2}{r}{86\%} & \\
    \hline
    Energy per Inference & \SI{13.86}{\micro\joule} & \multicolumn{2}{r}{\SI{3.2}{\micro\joule}}& & \multicolumn{2}{r}{\textbf{\SI{2.72}{\micro\joule}}} & \\
    \hline
    Core Area [$\text{mm}^2$] & 1.4 & \multicolumn{2}{r}{\textbf{0.39}}& &  \multicolumn{2}{r}{2.96} & \\
    \hline
    Voltage [V] & 0.65 & 0.37 & 0.75 & & 0.5 & 0.9 & \\
    \hline
    Throughput [TOp/s] & 2.8 & 3.4 & \textbf{163} & & 16 & 56 & \\
    \hline
    Peak Core Energy Efficiency [TOp/s/W]  & 230 & 617 & 269& & \textbf{1036} & 446 &\\
    \tabucline[1.25pt]{-}
  \end{tabu}
  \label{tab:comparison}
\end{table*}

Table~\ref{tab:comparison} shows a comparison of CUTIE on the Kraken \gls{soc} with state-of-the-art highly quantized digital convolutional network accelerators. CUTIE achieves a peak throughput of \SI{56}{TOp\per\second} and a peak energy efficiency of \SI{1036}{TOp\per\second\per\watt}, surpassing the highest reported efficiency in the literature: a \gls{bnn} accelerator manufactured in a more advanced technology node \cite{knag_617_2020}. \rev{As demonstrated in \cite{scherer_cutie_2022} by the use of post-layout simulation of a larger configuration of CUTIE, the high energy efficiency of CUTIE can be explained mainly by two design characteristics: The source of CUTIE's efficiency is the minimization of data movement, which limits the efficiency of comparable accelerators. This is achieved by the fully unrolled architecture, which minimizes the number of accesses to each data item. Secondly, the simple ternary processing elements and the use of very wide addition trees leverages sparsity in ternary data indirectly. This is shown in \cite{scherer_cutie_2022}, where ternarized networks with very sparse activations and weights reduce the inference energy cost on CUTIE by 36\%. }
\rev{In this work, we improve on these characteristics by optimizations in the front- and backend design flow, as well as using a smaller CUTIE configuration. }

To evaluate our \gls{tcn} extensions we compare our design with traditional \gls{tcn} accelerators, as well as with \gls{snn} accelerators, which purportedly are more energy efficient for sparse, event-based time-series data like \gls{dvs}. 

Although there are no standard benchmarks or datasets for \glspl{tcn}, we can compare the average energy efficiency over an inference for state-of-the-art designs. In \cite{giraldo_efficient_2021}, the authors propose a \gls{tcn} accelerator design for continuous, ultra-low-power keyword spotting. While running 64 inferences of a \SI{1.5}{MOp\per inference} network per second, they achieve an average power consumption between \SI{5}{\micro\watt} and \SI{15}{\micro\watt}, leading to average energy efficiency of \SI{6.4}{TOp\per\second\per\watt} to \SI{19.2}{TOp\per\second\per\watt}, measured by post-synthesis simulation. In direct comparison, our measured average energy cost per operation on the \gls{dvs} network is around 5 - 15$\times$ lower.

Even when comparing the performance of the \gls{tcn} extensions with state-of-the-art \gls{snn} accelerators on the DVS 128 dataset, our implementation meets the best reported accuracy using \glspl{tnn} in literature, a network deployed on the IBM Truenorth platform, which achieves a statistical accuracy of 94.6\%, just 0.1\% better than our ternary \gls{tcn}, while requiring 3250$\times$ more energy per inference \cite{amir_low_2017} than our design.

When comparing with a modern Intel \SI{14}{\nano\meter} accelerator implementation, the measured energy per inference of \SI{5.5}{\micro\joule} beats the best reported energy efficiency on a similar \gls{dvs} and EMG dataset, an \gls{snn} running on the Intel Loihi platform and achieving an accuracy of 96.0\%, by a factor of 63.4$\times$ \cite{ceolini_hand-gesture_2020}.
\section{Conclusion}
We presented the CUTIE implementation in the Kraken \gls{soc} and evaluated its performance. By exploiting minimized data movement and switching activity coupled with aggressive voltage scaling, we achieve a peak efficiency of \SI{1036}{TOp\per\second\per\watt}, surpassing the SoA in ultra-low-energy \gls{cnn} inference by a factor of $1.67 \times$.
Similarly, the implemented \gls{tcn} extensions are demonstrated to surpass the energy efficiency of the state-of-the-art by a factor of 5$\times$. 
\section*{Acknowledgement}
The authors would like to thank \textit{armasuisse Science \& Technology} for funding this research, and IniVation for kindly lending us a DVS Camera.
\bibliography{./main}

\begin{IEEEbiographynophoto}{Moritz Scherer} received the B.Sc. and M.Sc. degree in electrical engineering and information technology from ETH Zürich in 2018 and 2020, respectively, where he is currently pursuing a Ph.D. degree at the Integrated Systems Laboratory. Contact him at scheremo@iis.ee.ethz.ch.
\end{IEEEbiographynophoto}
\begin{IEEEbiographynophoto}{Alfio Di Mauro} received the M.Sc. degree in electronic engineering from the Electronics and Telecommunications Department (DET), Politecnico di Torino, in 2016. He is currently pursuing a Ph.D. degree with the Integrated System Laboratory (IIS), ETH Zürich. Contact him at adimauro@iis.ee.ethz.ch.
\end{IEEEbiographynophoto}
\begin{IEEEbiographynophoto}{Tim Fischer} received the B.Sc. and M.Sc. degree in electrical engineering and information technology from ETH Zürich in 2018 and 2021, respectively, where he is currently pursuing a Ph.D. degree at the Integrated Systems Laboratory. Contact him at fischeti@iis.ee.ethz.ch.
\end{IEEEbiographynophoto}
\begin{IEEEbiographynophoto}{Georg Rutishauser} received his B.Sc. and M.Sc.degrees in Electrical Engineering and Information Technology from ETH Zürich in 2015 and 2018, respectively. He is currently pursuing a Ph.D. degree at the Integrated Systems Laboratory at ETH Zürich. Contact him at georgr@iis.ee.ethz.ch.
\end{IEEEbiographynophoto}
\begin{IEEEbiographynophoto}{Luca Benini} holds the Chair of Digital Circuits and Systems at ETH Zürich and is a Full Professor with the Universita di Bologna. He is a Fellow of the ACM and the IEEE and a member of the Academia Europaea. Contact him at lbenini@iis.ee.ethz.ch.
\end{IEEEbiographynophoto}

\end{document}